\documentstyle[aps,preprint,prb]{revtex}
\begin{document}
\title{Spin Gaps in High Temperature Superconductors}
\author{A. J. Millis}
\address{AT\&T Bell Laboratories, 600 Mountain Ave, Murray Hill NJ
07974}
\author{L. B. Ioffe}
\address{Physics Department, Rutgers University, Piscataway, NJ 08855}
\author{Hartmut Monien}
\address{Theoretische Physik,
ETH Zurich,
CH-8093 Zurich,
Switzerland}
\maketitle
\begin{abstract}
The phenomenology and theory of spin gap effects in
high temperature superconductors is summarized. It is argued that
the spin gap behavior can only be explained by a model of charge 0
spin 1/2  fermions which become paired into singlets and that  there
are both theoretical and experimental reasons for
believing that the pairing is greatly enhanced in the
bilayer structure of the $YBa_2Cu_3O_{6+x}$ system.
\end{abstract}

\newpage

The spin dynamics of the high $T_c$ superconductors are anomalous
in several ways.
The aspect of
interest here is shown
in Fig. 1, in which the uniform spin susceptibility
$\chi_0 (T) \equiv lim_{q \rightarrow 0} \chi (q, \omega =0)$
is plotted for several high $T_c$ materials.
Focus first on the data for
$YBa_2Cu_4O_8$.  This is typical of the "underdoped" members of the
yttrium-barium family of high temperature superconductors.
Now the optimally doped member of this family,
$YBa_2Cu_3O_7$, displays the $\chi(T)$ expected for
a fermi liquid: it is large, about 3 states/eV-Cu \cite{Millis90}, and
temperature independent.
$YBa_2Cu_4O_8$ displays two regimes.
For $T > T^* \approx 200K$,
$\chi_0 (T) / \mu_B^2 \approx A+BT$ with $A = 1.5 states/eV-Cu$
and $B = 1.6 \times 10^{-3} states/eV-Cu-K$; for
$T < T^*$, $\chi_0 (T) $ drops more rapidly.
Neither temperature regime is compatible with
fermi liquid theory; the low temperature regime is particularly difficult to
explain.
The oxygen
relaxation rate $1/^{17}T_1$,
shown in Fig. 2 for several high $T_c$ materials,
displays similar anomalies.
This is related to the low $\omega$ limit of the imaginary part
of the spin susceptibility as discussed in \cite{Millis90}.
The data shown in Fig. 2 therefore imply that the
anomalous behavior found in $YBa_2Cu_4O_8$ at
$\omega =0$ and $q \rightarrow 0$ persists for a range of
$q$, at small $\omega$.

In a fermi liquid the  spin response is determined by
a continuum of particle-hole pair excitations.
The non-fermi liquid temperature dependences
observed in $YBa_2Cu_4O_8$ imply that in this
material the particle-hole continuum has been
modified in some essential way.
We argue that (i) the data for $T > T^*$
imply that in $YBa_2Cu_4O_8$  a particle-hole continuum
of spin excitations still exists,  (ii) the behavior at
$T<T^*$ can only be explained if the fermions making
up the particle-hole continuum are pairing into singlets,
(iii) because this singlet pairing is not
associated with the onset of superconductivity or superconducting fluctuations,
spin-charge separation must occur and (iv) there
are both theoretical and experimental reasons for
believing that the pairing is greatly enhanced in the
bilayer structure of the Y-Ba system.
These points are not new; here we aim to clarify the issues
and summarize the present  theoretical
understanding.

We consider $T >T^*$ first.  In this regime the copper NMR
relaxation rates $^{63}T_1^{-1}$ and $^{63}T_2^{-1}$,
both increase by almost a factor of 2 as T is lowered from
$400K$ to $200K$ \cite{Itoh92}. $T_1$ measures
 $\lim_{\omega \rightarrow 0} \sum_q$
$\chi^{\prime\prime} (q, \omega ) / \omega$ while $T_2$ measures
$(\sum_q \chi^{\prime} (q, \omega =0 ) ^2)^{1/2}$ \cite{Pennington91}.  The
simultaneous increase
can to our knowledge only be explained by models which assume proximity to a
$T=0$ magnetic critical point or phase.  NMR and neutron
scattering are not consistent (see e.g. \cite{Walstedt94});
our view is that
there are spin fluctuations not seen by neutron scattering,
but the issue is
not resolved.

Three classes of models have been proposed:
 ``fermi liquid''
models with overdamped spin excitations, ``$z=1$'' models with undamped or
weakly damped spin waves, and
``spin liquid'' models.
In fermi liquid models, the particle-hole continuum is modified by the
presence of overdamped spin excitations.
The spin excitations are overdamped because decay of a
spin excitation into a particle-hole pair is allowed.
There are two sub-cases --- either the magnetic wavevector is inside
the particle-hole continuum, or it is a ``$2p_F$'' wavevector
of the fermi surface.
The critical phenomena of these cases have been worked
out \cite{MillisQCrit,Chipaper}, and the comparison to
data is not favorable, as first pointed out in ref. \cite{Sokol93}
The principle difficulty concerns the Cu
$T_2$ rate, which is predicted to vary with temperature as $(T_1T)^{-1/2}$, if
the ordering wavevector is inside the continuum, or more weakly,
if $Q=2p_F$,
whereas the data suggest $T_2^{-1} \propto 1/T_1T$ \cite{Walstedt94,Sokol93}.

The $z=1$ models take as their starting point the
magnetic insulating parent compounds, which are antiferromagnets
with spin waves of dispersion
$\omega = c[|\vec{k}-\vec{Q}|]$
$( \vec{Q} = ( \pi , \pi )$ is the ordering vector
and in $La_2CuO_4$ $\hbar c \sim 0.75 eV- \AA$).
It is assumed that the doping which converts these materials
to superconductors also  changes the dispersion to
$\omega^2 = c^2 [(\vec{k}-\vec{Q})^2 + \Delta^2]$.
The critical properties associated with opening the gap
have been extensively studied \cite{Chakravarty}.
For $T> \Delta > 0$, the copper NMR $T_1$ and $T_2$ rates are
predicted to obey
$T_2/T_1T \sim$ constant $\sim c$ in agreement
with data \cite{Itoh92,Sokol93}.
However, the very restrictive kinematics of spin waves implies
that e.g. the oxygen relaxation rate in the spin-wave-only model
is given by $1/^{17}T_1T \sim T^2$ \cite{MillisMonien94}.
It is therefore argued \cite{Sokol93} that the model must be supplemented
by a particle hole continuum of fermions which couple only weakly
to the spin waves but contribute to the small $q$ response, so that
$\chi_0 (T) = \chi_{fermion} + \chi_{spin wave}$.
However, a quantitative comparison of the theory to
$YBa_2Cu_4O_8$ \cite{MillisMonien94}
found that  the observed $T_2 / T_1 T$
ratio implied
$c \cong 0.35 \; eV - \AA$, implying
$d\chi_{spin wave}/dT= 1 \times 10^{-2} states/ev-Cu-K$,
${\it six}$ times greater than the observed $d\chi/dT$.

A third alternative is the ``spin liquid'' picture.
This is based on a gauge theory formalism
which is one implementation of Anderson's
spin-change separation hypothesis \cite{Ioffe-Larkin}.
The spin response is given by a particle-hole
continuum of spin $1/2$
charge 0 fermionic "spinons" which interact via a
gauge field.
It has been shown  that the gauge field interaction may lead to
divergences in the ``$2p_F$'' susceptibilities
of the spinons even if the interaction is not tuned to a critical value,
and parameters exist for which the
physical consequences are roughly
consistent with available data \cite{AIM}.

We now turn to  point (ii).
For $T < T^* \approx 200K$,
$\chi_0 (T)$ drops below the $\chi_s = A+BT$ characterizing
the higher $T$ behavior.
For $100K < T < 200K$, $\chi_0(T)$ is roughly linear
in $T$ and extrapolates to a slightly negative value at $T=0$.
In other words, in this regime {\it one should
think of the $\chi_0(T)$ as being due to thermal excitations above a ground
state with a vanishing spin susceptibility}.
In order to have $\chi_0(T)=0$ at $T=0$ in
a spin rotation invariant system one must have a ground state which
is a singlet with a gap to spin excitations.
We have already argued that there is a
particle-hole continuum of spin excitations in $YBa_2Cu_4O_8$.
The only known method of opening a gap in a
particle-hole continuum is to pair the fermions into singlets,
as occurs in a conventional superconductor.
Note in particular that the  effect of actual or incipient
antiferromagnet  order on the spin
susceptibility is known \cite{Chipaper}
not to produce a gap; instead it leads in two spatial dimensions to
$\chi_0 (T) = \chi_0 + DT$, with $\chi_0 > 0$.

The pairing is unlikely
to be due to conventional superconductivity for two reasons:
the scale, $T^* \approx 200K$,
is much larger than the largest superconducting $T_c$,
observed in any member of the Y-Ba family,
and the observed resistivity is very
different from the ``paraconductivity'' observed in materials
with strong superconducting fluctuations.
In particular, $YBa_2Cu_3O_{6.7}$ exhibits spin
gap effects very similar to those found in
$YBa_2Cu_4O_8$ but its resistivity
has upward curvature for $80K < T <150K$ and is rather
flat between 100K and $T_c=60K$ \cite{resistivity}.
If strong superconducting fluctuations
were present, the resistivity would drop rapidly in this regime.
Thus we believe that theories based on formation of conventional
Cooper pairs via an attractive interaction \cite{Randeria92} are unlikely to be
be relevant.  Rather, the correct theory must involve spin-charge separation
in some form.

Next, we argue
that although anomalous temperature dependences are present,
there is no strong evidence for
singlet pairing in $La_{2-x}Sr_xCuO_4$, where
interplane coupling is believed to be weak.
Spin susceptibility data obtained as in \cite{Millis93} but
using more reliable  data of \cite{Nakano94} are shown in Fig. 1;
although there is a pronounced downturn in the susceptibility for
$x=0.08$ and $x=0.14$ it is also clear that
$lim_{T \rightarrow 0} \chi_s (T) > 0$ in
these materials.
Similarly, $^{17}T_1T^{-1}$, plotted in Fig. 2 at
$x=0.14$, shows no sign of a downturn.
Finally, the Cu relaxation rates $1/T_1T$ in
$La_{2-x}Sr_xCuO_4$ are monotonic in $T$ \cite{Walstedt94}, in contrast
to the Cu relaxation rate in $YBa_2Cu_4O_8$ which exhibits
a pronounced downturn beginning almost a factor of two above $T_c$
\cite{Itoh92}.  We conclude  that
the bilayer structure of $YBa_2Cu_4O_8$ is important. This conclusion is not
universally accepted; for an alternative point of view see \cite{Sokol93}.

There is in fact evidence that the two planes in a bilayer are magnetically
coupled.  Neutron scattering experiments have only detected fluctuations
in which the spins on one plane are perfectly anticorrelated
with spins on the other \cite{Tranquada92}.  An NMR $T_2$
"interplane relaxation"
experiment in which
Cu spins on one plane are pumped and spins on the adjacent plane
are probed has been proposed by Monien and Rice \cite{Monien95}
and performed by Stern et. al. \cite{Stern95}.  The interplane rate is found
to be large (of the order of the in-plane $T_2$) and rapidly temperature
dependent.

Theories of the spin gap have been discussed by many authors
\cite{Tanamoto92,Altshuler92,JETP,Ubbens94,Strong95}.  Because
one must deal with "spinons", the effects of  gauge interactions must be
included.  These lead to a large inelastic lifetime and thus strong
pairbreaking which in general suppresses \cite{Ubbens94} the spinon-pairing
instabilities found  \cite{Tanamoto92} in one-plane models (for
an exception, see \cite{JETP}, where in one model a very weak
instability is found).  However, in two-plane models the enhanced
in-plane magnetic susceptibility leads \cite{JETP,Ubbens94} to a large
enhancement of the
between-planes pairing originally proposed in  \cite{Altshuler92}.
In fact, the pairing kernel has essentially the same singularity as
the theoretical expression for the "interplane relaxation"
$T_2$ discussed above,
so this experiment may be regarded as evidence that the singular
interaction required by the interplane pairing theories exists and has the
correct order of magnitude.

To conclude:  a magnetic susceptibility which decreases as the temperature is
decreased is observed in underdoped high temperature superconductors
and is difficult to explain within fermi liquid theory.  We have argued that
one should distinguish two regimes:  a $\chi_0(T) \approx A + BT$ regime
which occurs in many materials and must be understood
in terms of antiferromagnetism and a particle-hole continuum,
and a "spin-gap" regime in which $\chi$ drops rapidly to
zero, which occurs only in the yttrium-barium family and  must
be understood in terms of pairing of chargeless fermionic "spinons".

We thank Bertram Batlogg for very helpful discussions of susceptibility
measurements.

\newpage
\begin{figure}
\caption{Spin susceptibilities for $YBa_2Cu_4O_8$ (filled squares)
 and $La_{2-x}Sr_xCuO_4$
(x=0.08, open circles, x=0.14 open triangles, x=0.18, open squares).
The solid line is the slope predicted by the quantum critical regime of the z=1
theory with the spin wave velocity  appropriate
to $La_2CuO_4$.}
\end{figure}

\begin{figure}
\caption{Oxygen relaxation rates from ref [1] (Y-Ba) and [4] (La-Sr).}
\end{figure}

\begin{thebibliography}{99}
\bibitem{resistivity}
See, e.g. Fig 3 of B. Batlogg, p. 37 in ref [1]. This figure
presents data for $YBa_2Cu_3O_{6.7}$ which has spin gap behavior
similar to $YBa_2Cu_4O_8$.  The resistivity flattening is present
in only the former material, which has more impurity scattering.  The
sensitivity of $\rho$ to impurities shows that there is no paraconductivity.
\end{thebibliography}
\end{document}